\documentclass[parindent=0pt]{article}
\usepackage{booktabs}
\usepackage{makecell}
\usepackage{subcaption}

\usepackage[utf8]{inputenc}     % for éô
\usepackage[english]{babel}     % for proper word breaking at line ends
\usepackage[a4paper, left=1in, right=1in, top=1.25in, bottom=1.25in]{geometry}
\usepackage{graphicx}           % for ?

\usepackage{amsmath,amssymb}    % for better equations
\usepackage{amsthm}             % for better theorem styles
\usepackage{mathtools}          % for greek math symbol formatting
\usepackage{enumitem}           % for control of 'enumerate' numbering
\usepackage{listings}           % for control of 'itemize' spacing
\usepackage{todonotes}          % for clear TODO notes
\usepackage{hyperref}           % page numbers and '\ref's become clickable
\usepackage{float}
\usepackage{amsthm}
\usepackage{amsmath}
\usepackage{amscd}
\usepackage{amssymb}
\usepackage{mathrsfs}
\usepackage{pdfpages}
\usepackage{setspace}
\usepackage{physics}
\usepackage{tcolorbox}

\usepackage{xcolor}
\usepackage{tikz}
\usepackage[toc,page,title,titletoc,header]{appendix}
 \usepackage{url}
 \usepackage{babel} 
\usepackage{csquotes}
\usepackage[f]{esvect}
\usepackage{blindtext}

\newcommand{\C}{\mathbb{C}}

\newcommand{\g}{\mathfrak{g}}

\newcommand{\diag}{\mbox{diag}}

\let\phi=\varphi
\let\epsilon=\varepsilon

\theoremstyle{definition}
\newtheorem{result}{Result}

\newtheorem*{definition*}{\textit{Definition}}
\newtheorem*{proposition*}{\textit{Proposition}}
\newtheorem*{conjecture*}{\textit{Conjecture}}

\newcommand{\adots}{\mathinner{\mkern1mu \raise1pt
\hbox{.} \mkern2mu \raise4pt \hbox{.} \mkern2mu \raise7pt
\vbox{\kern7pt \hbox{.}} \mkern1mu}}
\def\[{[\![}
\def\]{]\!]}

\newlist{regimes}{enumerate}{1}
\setlist[regimes]{label={\textbullet\ Regime (\arabic*):}, leftmargin=*}
\usepackage{authblk} % Pour gérer les affiliations multiples

\title{%
   The eigenvalues and eigenvectors of finite-rank normal perturbations of large rotationally invariant non-Hermitian matrices
}

  \author[1,2]{Pierre Bousseyroux\thanks{Email: pierre.bousseyroux@polytechnique.edu}}
\author[3]{Marc Potters}

\affil[1]{Econophysics Lab, Institut Louis Bachelier, 28 Pl. de la Bourse, Palais Brongniart, 75002 Paris, France}
\affil[2]{LadHyX, UMR CNRS 7646, Ecole Polytechnique, Institut Polytechnique de Paris, 91128 Palaiseau, France}
\affil[3]{Capital Fund Management, Paris, France}

\newenvironment{remarks}{
  \par\vspace{1ex}
  \noindent\textbf{Remarks.}\begin{itemize}\setlength\itemsep{0pt}}
  {\end{itemize}\par\vspace{1ex}}

\makeatletter
\renewcommand{\@fnsymbol}[1]{%
  \ifcase#1\or
    \ensuremath{\dagger}\or
    \ensuremath{\ddagger}\or
    \ensuremath{\mathsection}\or
    \ensuremath{\mathparagraph}\or
    \ensuremath{\|}\or
    \ensuremath{\dagger\dagger}\or
    \ensuremath{\ddagger\ddagger}\or
    \ensuremath{\mathsection\mathsection}\or
    \ensuremath{\mathparagraph\mathparagraph}\else
    \@ctrerr
  \fi}
\makeatother

\begin{document}
\maketitle

\begin{abstract}
We study finite-rank normal deformations of rotationally invariant non-Hermitian random matrices. Extending the classical Baik-Ben Arous-Péché (BBP) framework, we characterize the emergence and fluctuations of outlier eigenvalues in models of the form \(\vb{A} + \vb{T}\), where \(\vb{A}\) is a large rotationally invariant non-Hermitian random matrix and \(\vb{T}\) is a finite-rank normal perturbation. We also describe the corresponding eigenvector behavior. Our results provide a unified framework encompassing both Hermitian and non-Hermitian settings, thereby generalizing several known cases.
\end{abstract}

In high-dimensional random matrix theory, a low-rank deterministic perturbation can significantly alter the extreme eigenvalues of a large random matrix. For a random matrix \(\vb{A}\) with a limiting spectral distribution supported on a bounded interval (the bulk), adding a finite-rank matrix \(\vb{T}\) yields \(\vb{M} := \vb{A} + \vb
{T}\), whose eigenvalues mostly remain within the bulk. However, sufficiently strong perturbations can cause a few eigenvalues to detach and form outliers. This separation marks the Baik–Ben~Arous–Péché (BBP) phase transition~\cite{BBP2005,Peche2006}, first identified in the context of spiked covariance models.

For Hermitian random matrices, the theory of finite-rank deformations is well established. More generally, for any rotationally invariant Hermitian \(\vb{A}\)\footnote{In the sense that $\vb{A}$ and $\vb{U}\vb{A}\vb{U}^*$ have the same distribution for every unitary matrix $\vb{U}$.} with a compact limiting eigenvalue distribution, a Hermitian finite-rank perturbation~$\vb{T}$ can generate an outlier of \(\vb{M} = \vb{A}+ \vb{T}\) converging to a deterministic limit outside the original bulk~\cite{Peche2006,FP2007,Capitaine2009,KnowlesYin2014}. This transition is accompanied by changes in both eigenvalue fluctuations and eigenvector structure: in the subcritical regime, the largest eigenvalue of \(\vb{M}\) of size $N$ fluctuates as in the unperturbed case (e.g. Tracy–Widom fluctuations on an \(N^{2/3}\) scale), whereas a well-separated supercritical outlier exhibits smaller \(O(N^{-1/2})\) Gaussian fluctuations~\cite{BBP2005,Capitaine2012,BaiYao2008}. Likewise, below the threshold the leading eigenvector of \(\vb{M}\) is asymptotically orthogonal to the spike, while above the threshold it develops a nonzero alignment with the spike direction~\cite{Paul2007,benaych2011eigenvalues}. A related transition for the eigenvalues only, in the case of complex symmetric finite-rank perturbations, has also been investigated in~\cite{Dubach2023}.

Analogous outlier phenomena occur in non-Hermitian random matrices with low-rank perturbations. For instance, if~$\vb{A}$ is an $N\times N$ noise matrix with i.i.d.\ entries (such as a complex Ginibre matrix, whose eigenvalues are asymptotically uniform in a disk~\cite{Girko1985_CircularLaw_Eng}), adding a finite-rank normal perturbation~$\vb{T}$\footnote{An operator~$\vb{T}$ is said to be \emph{normal} if it commutes with its adjoint, i.e., $\vb{T}\vb{T}^* = \vb{T}^*\vb{T}$, where $\vb{T}^* := \overline{\vb{T}}^{\,T}$ denotes the conjugate transpose of~$\vb{T}$.} can produce outliers outside the support of~$\vb{A}$’s spectrum~\cite{Tao2013}. This criterion has been extended to other non-Hermitian ensembles, such as elliptically distributed random matrices~\cite{ORourke2014}, more general deformed i.i.d.\ models~\cite{bordenave2016outlier}, and bi-invariant ensembles\footnote{We say that a random matrix~$\vb{M}$ is bi-invariant if $\vb{M}$ and $\vb{U}\vb{M}\vb{V}$ have the same distribution for all unitary matrices~$\vb{U}$ and~$\vb{V}$.}~\cite{Benaych2016}. In all these settings, the outlier eigenvalues exhibit Gaussian fluctuations on an~$O(N^{-1/2})$ scale when well separated from the bulk~\cite{bordenave2016outlier,Benaych2016}.
The associated eigenvectors, however, have not yet been characterized in a comparably general framework for rotationally invariant non-Hermitian ensembles. In the specific case of complex Ginibre matrices with a fixed additive rank-one deformation, Fyodorov recently derived a finite-$N$ formula for the joint density of an eigenvalue and its associated right eigenvector, and analyzed the large-$N$ limit~\cite{fyodorov2025kac}.

The aim of this paper is to provide a general framework for studying outliers, their associated eigenvectors, and the corresponding fluctuations for matrices of the form \(\vb{A} + \vb{T}\), where \(\vb{T}\) is a finite-rank normal matrix and \(\vb{A}\) is a (possibly non-Hermitian) rotationally invariant random matrix. The results presented here extend previous analyses and rely extensively on the theory developed in~\cite{bousseyroux1}.

\section{General setting and main result}

We now aim to generalize all previously known results by encompassing them within the following general framework, where $\vb{A}$ denotes a large, rotationally invariant random matrix, not necessarily Hermitian, such that $\vb{U}\vb{A}\vb{U}^*$ has the same distribution as $\vb{A}$ for any unitary matrix $\vb{U}$. Let $\vb{T}$ be a large finite-rank perturbation, meaning that there exist an integer $n$ and complex scalars $s_1, \ldots, s_n \in \C$, as well as unit vectors $u_1, \ldots, u_n \in \C^N$, such that
\begin{equation}
    \vb{T} = \sum_{k=1}^n s_k\,u_k\,u_k^*,
\end{equation}
where $u^* := \overline{u}^{\,T}$ for any $u \in \C^N$. The main analytical tool will be the Stieltjes transform $\g_{\vb{A}}$ of $\vb{A}$, defined in the large-$N$ limit as  
\begin{equation}
    \g_{\vb{A}}(z)
    \;:=\;
    \lim_{N\to+\infty} \frac{1}{N}\Tr\!\bigl[(z\vb{1} - \vb{A})^{-1}\bigr],
    \qquad z \in \C.
\end{equation}
We will also make use of the \(\mathcal{R}_{1,\vb{A}}\) and \(\mathcal{R}_{2,\vb{A}}\) transforms associated with~\(\vb{A}\), as well as their corresponding multivalued functions \(\widetilde{\mathcal{R}}_{1,\vb{A}}\) and \(\widetilde{\mathcal{R}}_{2,\vb{A}}\), introduced in~\cite{bousseyroux1}. We refer the reader to that reference, since the proofs presented here rely extensively on it. In particular, recall Result~4, which states that if \(z\) lies outside the support of the limiting continuous spectrum of~\(\vb{A}\), and if \(\g_{\vb{A}}(z)\) is not identically equal to \(0\), then one can find appropriate branches \(\mathcal{R}_{1,\vb{A}}\) and \(\mathcal{R}_{2,\vb{A}}\) such that
\begin{equation}\label{eq1}
    \partial_\alpha \mathcal{R}_{1,\vb{A}}(0, \g_{\vb{A}}(z))
    =
    \frac{1}{|\g_{\vb{A}}(z)|^2}
    -
    \frac{1}{h_{\vb{A}}(z)},
\end{equation}
and
\begin{equation}\label{eq2}
    \g_{\vb{A}}(z)
    =
    \frac{1}{z - \mathcal{R}_{2,\vb{A}}(0, \g_{\vb{A}}(z))},
\end{equation}
where
\begin{equation}
    h_{\vb{A}}(z)
    =
    \lim_{N\to +\infty}
    \frac{1}{N}
    \tr\!\left(
    \left[
    (z\vb{1} - \vb{A})(z\vb{1} - \vb{A})^*
    \right]^{-1}
    \right).
\end{equation}

We can now state the main result, which consists of three parts, whose proofs follow essentially the same steps as in the Hermitian case and are given respectively in Appendices~\ref{sec:outlier}, \ref{sec:eigenvectors}, and~\ref{sec:fluctuations}. For simplicity, we restrict ourselves to the case~$n=1$, and the notations used in the proofs are introduced at the beginning of the Appendices.
\begin{result}
Fix $1 \le k \le n$.  
In the high-dimensional limit, the rank-one perturbation $s_k\,u_k\,u_k^*$ may produce potential outliers of $\vb{M}:=\vb{A} + \vb{T}$, which are solutions of
\begin{equation}\label{solution_outlier}
    \g_{\vb{A}}(z) = \frac{1}{s_k}.
\end{equation}

\begin{enumerate}
    \item[\textbf{(i)}] \textbf{Outlier location.}  
    For each such outlier $z$, one can equivalently write
    \begin{equation}
        z
        \;=\;
        s_k + \mathcal{R}_{2, \vb{A}}\!\left(0, \frac{1}{s_k}\right),
    \end{equation}
    where $\mathcal{R}_{2, \vb{A}}$ is a determination of the multivalued function $\widetilde{\mathcal{R}_{2, \vb{A}}}$.

    \item[\textbf{(ii)}] \textbf{Eigenvector alignment.}  
    Let $\phi_k$ denote a normalized eigenvector associated with the outlier $z$. Then,
    \begin{equation}\label{overlap}
        \bigl|\langle u_k , \phi_k \rangle\bigr|^2
        \;\xrightarrow[N \to +\infty]{}\;
        1 - \frac{\partial_{\alpha}\mathcal{R}_{1, \vb{A}}\!\left(0, \frac{1}{s_k}\right)}{|s_k|^2},
    \end{equation}
    where $\mathcal{R}_{1, \vb{A}}$ is a determination of $\widetilde{\mathcal{R}_{1, \vb{A}}}$.

    \item[\textbf{(iii)}] \textbf{Fluctuations.}  
    For large~$N$, the fluctuations of the outlier position~$z$ are Gaussian with variance~$\frac{\sigma^2}{N}$, where
\begin{equation}\label{express_variance}
    \sigma^2 = 
    \frac{\partial_{\alpha}\mathcal{R}_{1,\vb{A}}\!\left(0, \frac{1}{s_k}\right) 
    \bigl|\,s_k^2 - \partial_{\beta}\mathcal{R}_{2,\vb{A}}\!\left(0, \frac{1}{s_k}\right)\bigr|^2}
    {|s_k|^4 - \partial_{\alpha}\mathcal{R}_{1,\vb{A}}\!\left(0, \frac{1}{s_k}\right)\,|s_k|^2}.
\end{equation}
where $\mathcal{R}_{1,\vb{A}}$ denotes a determination of $\widetilde{\mathcal{R}_{1,\vb{A}}}$, and $\mathcal{R}_{2,\vb{A}}$ denotes a determination of $\widetilde{\mathcal{R}_{2,\vb{A}}}$.

\end{enumerate}
\end{result}

\begin{remarks}
    \item Using formula~\eqref{overlap}, we conjecture that the transition occurs when a point $s_k \in \mathbb{C}$ crosses the curve defined by
\begin{equation}
1 - \frac{\partial_{\alpha}\mathcal{R}_{1, \vb{A}}\!\left(0, \frac{1}{s_k}\right)}{|s_k|^2} = 0,
\end{equation}
where $\mathcal{R}_{1,\vb{A}}$ denotes a determination of $\widetilde{\mathcal{R}}_{1,\vb{A}}$.
This provides a natural extension of the BBP transition threshold to the complex plane.

        \item In the Hermitian setting, i.e. when both $\vb{A}$ and $\vb{T}$ are Hermitian (and hence the $s_k$ are real), the theory is well known (see~\cite{benaych2011eigenvalues,potters2020first}).  
    Looking at equations~\eqref{eq1} and~\eqref{eq2}, one notes that $x \mapsto \mathcal{R}_{2, \vb{A}}(0, x)$ is simply the usual $R$-transform of $\vb{A}$, and
    \begin{equation}
        \partial_{\alpha} \mathcal{R}_{1, \vb{A}}(0, x) = \frac{\Im(R(x))}{\Im(x)},
    \end{equation}
    which converges to $R'(x)$ as $\Im(x)\to 0$.
    We then recover the classical expressions for outliers:
    \begin{equation}
        s_k + R\!\left(\frac{1}{s_k}\right),
    \end{equation}
    and for the eigenvector overlap:
\begin{equation}
    1 - \frac{R'\!\left(\frac{1}{s_k}\right)}{s_k^2}.
\end{equation}
The fluctuations then become
\begin{equation}
    R'\!\left(\frac{1}{s_k}\right)
    \!\left(1 - \frac{R'\!\left(\frac{1}{s_k}\right)}{s_k^2}\right).
\end{equation}
Our framework allows us to generalize the previous setting to the case where the $s_k$ become complex, while the eigenvalues of $\vb{A}$ remain real. In particular, one finds that outliers become visible when they cross the curve in the complex plane defined by
\begin{equation}\label{transition}
    \Im\!\left(z + R\!\left(\frac{1}{z}\right)\right) = 0,
\end{equation}
which generalizes the usual BBP threshold. In the case where $\vb{A}$ is a Ginibre matrix, equation~\eqref{transition} reduces to the unit circle, a result already known from~\cite{liu2025phase}.

    \item We chose to express the result using the different possible determinations of $\mathcal{R}_{1, \vb{A}}$ and $\mathcal{R}_{2, \vb{A}}$. This simply provides a convenient way to describe the multiple solutions of equation~\eqref{solution_outlier} while keeping the connection with~\eqref{eq2}.  
    Note that the number of outliers created by a rank-one perturbation $s_k u_k u_k^*$ may be strictly greater than one and is bounded by the number of determinations.  
    This phenomenon is already known in the standard Hermitian case, where a rank-one perturbation can produce two outliers. For instance, consider a random matrix of the form
    \(\vb{A} := \vb{U}^* \diag(1,\dots,1,-1,\dots,-1)\vb{U}\),
    where half of the eigenvalues are $1$ and the other half are $-1$, and $\vb{U}$ is Haar distributed.  
    As shown in~\cite{potters2020first}, the $R$-transform is then given by
    \begin{equation}\label{exprR}
        R(g) = \frac{-1 + \sqrt{1 + 4g^2}}{2g}.
    \end{equation}
    If one adds to $\vb{A}$ a rank-one Hermitian perturbation of the form $\sigma u u^*$, then two outliers may appear, given by $\sigma + R(1/\sigma)$, where $R$ runs over the two branches of~\eqref{exprR}. This reasoning also appears in \cite{benaych2011eigenvalues} and \cite{Belinschi2017}, although the authors do not use $R$-transforms explicitly, relying instead on equation~\eqref{solution_outlier}.

    \item Let us now consider the case where $\vb{A}$ is a large bi-invariant matrix. Using the single ring theorem~\cite{guionnet2011single}, we know that the support of the limiting spectral distribution of $\vb{A}$ is a ring, with inner and outer radii denoted by $r_{-, \vb{A}}$ and $r_{+, \vb{A}}$, respectively.  
    Furthermore, by elementary electrostatics and the rotational invariance of the distribution in the complex plane, one can show that
    \begin{equation}
    \g_{\vb{A}}(z) =
    \begin{cases}
        0, & \text{if } |z| < r_{\vb{A}}^{-}, \\
        \frac{1}{z}, & \text{if } |z| > r_{\vb{A}}^{+}.
    \end{cases}
    \end{equation}
    It follows that there is a unique outlier if and only if $|s_k| > r_{+, \vb{A}}$, which is simply given by $s_k$.  
    The variance of its fluctuations is
    \begin{equation}
        \frac{r_{+, \vb{A}}^2 |s_k|^2}{|s_k|^2 - r_{+, \vb{A}}^2}.
    \end{equation}
    This recovers the results presented in~\cite{Benaych2016, fyodorov2025kac}.  
    A new result obtained here concerns the associated eigenvectors:
    \begin{equation}
        \bigl|\langle u_k , \phi_k \rangle\bigr|^2
        \;\xrightarrow[N \to +\infty]{}\;
        1 - \frac{r_{+, \vb{A}}^2}{|s_k|^2},
    \end{equation}
    where $\phi_k$ denotes an eigenvector associated with the outlier $s_k$. This formula was previously obtained in the special case where $\vb{A}$ is a complex Ginibre matrix in~\cite{fyodorov2025kac}.
\end{remarks}

\paragraph*{Acknowledgements.} 

We are grateful to Florent Benaych-Georges and Jean-Philippe Bouchaud for their valuable insights. This research was conducted within the Econophysics \& Complex Systems Research Chair, under the aegis of the Fondation du Risque, the Fondation de l’Ecole polytechnique, the Ecole polytechnique, and Capital Fund Management.

\bibliographystyle{plain}
\bibliography{References.bib}

\appendix

\newpage
\section{APPENDICES}

We consider a large rotationally invariant random matrix~$\vb{A}$, and set $\vb{M} = \vb{A} + s\,u u^*$, where $s \in \C$ and $u \in \C^N$ is a unit vector. 

For any matrix~$\vb{B}$, we denote by $\vb{G}_{\vb{B}}(z)$ the resolvent of~$\vb{B}$ evaluated at a point $z \in \C$, defined by
\begin{equation}
    \vb{G}_{\vb{B}}(z) = (z\vb{1} - \vb{B})^{-1},
\end{equation}
where $\vb{1}$ denotes the identity matrix.

\subsection{Proof of (i)}\label{sec:outlier}

The Sherman–Morrison formula~\cite{sherman1950adjustment} yields
\begin{equation}\label{eq}
    \vb{G}_{\vb{M}}(z) 
    = 
    \vb{G}_{\vb{A}}(z) 
    + 
    s\,\frac{\vb{G}_{\vb{A}}(z)u u^*\vb{G}_{\vb{A}}(z)}{1 - s\,\bra{u}\vb{G}_{\vb{A}}(z)\ket{u}}.
\end{equation}
The rotational invariance of~$\vb{A}$ implies that
\begin{equation}
    \bra{u}\vb{G}_{\vb{A}}(z)\ket{u} 
    \xrightarrow[N\to\infty]{} 
    \g_{\vb{A}}(z).
\end{equation}
Therefore, in the high-dimensional limit, the eigenvalues of~$\vb{M}$ differ from those of~$\vb{A}$ if and only if the equation
\begin{equation}
    \g_{\vb{A}}(z) = \frac{1}{s}
\end{equation}
admits a solution. Using~\eqref{eq2}, we then obtain the first statement of the result.

\subsection{Proof of (ii)}\label{sec:eigenvectors}
Let $z$ be a possible outlier of~$\vb{M}$ and $\phi$ a unit eigenvector associated with it.  
We have
\begin{equation}
    (\vb{A} + s\,u\,u^*)\phi = z\,\phi,
\end{equation}
and thus
\begin{equation}
    (\vb{A} - z\vb{1})\phi = -s\,\langle u, \phi \rangle\,u.
\end{equation}
It follows that $\phi$ is proportional to
\begin{equation}
    (z\vb{1} - \vb{A})^{-1}u.
\end{equation}
Since $\phi$ is a unit vector, we have
\begin{equation}
    \phi 
    = 
    \frac{(z\vb{1} - \vb{A})^{-1}u}
    {\sqrt{\langle u \mid [(z\vb{1} - \vb{A})^*(z\vb{1} - \vb{A})]^{-1}u \rangle}}.
\end{equation}
Hence,
\begin{equation}
    |\langle u \mid \phi \rangle|^2 
    = 
    \frac{|\langle u \mid (z\vb{1} - \vb{A})^{-1} \mid u \rangle|^2}
    {\langle u \mid [(z\vb{1} - \vb{A})^*(z\vb{1} - \vb{A})]^{-1} \mid u \rangle}.
\end{equation}
Using the rotational invariance of~$\vb{A}$, this can be written as
\begin{equation}
    \frac{|\g_{\vb{A}}(z)|^2}{h_{\vb{A}}(z)}
\end{equation}
in the high-dimensional limit.
Finally, using~\eqref{eq1} together with~\eqref{solution_outlier}, we obtain the desired result~\eqref{overlap}.

\subsection{Proof of (iii)}\label{sec:fluctuations}
First, let $\vb{B}$ be a large rotationally invariant non-Hermitian matrix.  
Using Eq.~(32) from~\cite{bousseyroux1}, one obtains in particular that the fluctuations of a diagonal entry of $\vb{B}$ are Gaussian with variance
\begin{equation}\label{variance}
    \frac{\tau(\vb{B}\vb{B}^*) - |\tau(\vb{B})|^2}{N}
\end{equation}
as $N \to +\infty$, where $\tau:=\lim_{N\to +\infty} \Tr/N$ denotes the normalized trace.
The idea is to apply this result to a suitably chosen matrix. Returning to equation~\eqref{eq}, we have seen that for finite $N$, the possible outliers are given by the solutions of
\begin{equation}
    \bra{u}\vb{G}_{\vb{A}}(z)\ket{u} = \frac{1}{s},
\end{equation}
and that, as $N \to +\infty$, $z \to z^*$ where
\begin{equation}
    \g_{\vb{A}}(z^*) = \frac{1}{s}.
\end{equation}

Expanding $z$ around $z^*$, one finds that the fluctuations of $z$ have the same nature as those of $\bra{\psi}\vb{G}_{\vb{A}}(z)\ket{\psi}$ in the large-$N$ limit.  
In particular, using~\eqref{variance}, we obtain that these fluctuations are Gaussian with variance
\begin{equation}
    \frac{\tau(\vb{G}_{\vb{A}}(z^*)\vb{G}_{\vb{A}}(z^*)^*) - |\tau(\vb{G}_{\vb{A}}(z^*))|^2}{N\,\tau(\vb{G}_{\vb{A}}(z^*)^2)}.
\end{equation}
Finally, using equations~\eqref{eq1} and~\eqref{eq2}, one recovers~\eqref{express_variance}.

\end{document}